\documentclass{CHEP}
\usepackage{svg}

\usepackage[varg]{txfonts}   
\usepackage{booktabs}%
\usepackage{caption, subcaption}

\begin{document}
\title{Optimisation of ATLAS computing resource usage through a modern HEP Benchmark Suite via HammerCloud and Big PanDA}
%
%

\author{\firstname{Natalia Diana} \lastname{Szczepanek}\inst{1}\fnsep\thanks{\email{natalia.diana.szczepanek@cern.ch}} \and
        \firstname{Domenico} \lastname{Giordano}\inst{1}\fnsep\thanks{\email{domenico.giordano@cern.ch}} \and
        \firstname{Ivan} \lastname{Glushkov}\inst{2} \and
        \firstname{Gonzalo} \lastname{Menendez Borge}\inst{1} \and
        \firstname{Alessandro} \lastname{Di Girolamo}\inst{1} \and
        \firstname{Alexander} \lastname{Lory}\inst{3} \and
        \firstname{Ilija} \lastname{Vukotic}\inst{4}
}

\institute{CERN, European Laboratory for Particle Physics, Geneva, Switzerland 
\and
           University of Texas, Arlington, United States
\and
           Ludwig-Maximilians-Universit\"at, Munich, Germany
\and
           University of Chicago, United States
          }

\abstract{%
  In April 2023, HEPScore23, the new benchmark based on HEP specific applications, was adopted by WLCG, replacing HEP-SPEC06. As part of the transition to the new benchmark, the CPU corepower published by the sites needed to be compared with the effective power observed while running ATLAS workloads. One aim was to verify the conversion rate between the scores of the old and the new benchmark. The other objective was to understand how the HEPScore performs when run on multi-core job slots, so exactly like the computing sites are being used in the production environment. Our study leverages the HammerCloud infrastructure and the PanDA Workload Management System to collect a large benchmark statistic across 136 computing sites using an enhanced HEP Benchmark Suite. It allows us to collect not only performance metrics, but, thanks to plugins, it also collects information such as machine load, memory usage and other user-defined metrics during the execution and stores it in an OpenSearch database. These extensive tests allow for an in-depth analysis of the actual, versus declared computing capabilities of these sites. The results provide valuable insights into the real-world performance of computing resources pledged to ATLAS, identifying areas for improvement while spotlighting sites that underperform or exceed expectations. Moreover, this helps to ensure efficient operational practices across sites. The collected metrics allowed us to detect and fix configuration issues and therefore improve the experienced performance.
}
\maketitle
{\let\thefootnote\relax\footnote{\hspace{-5.8mm} Copyright 2025 CERN for the benefit of the ATLAS Collaboration. \\ Reproduction of this article or parts of it is
allowed as specified in the CC-BY-4.0 license.}}

\section{Introduction}
\label{section:intro}
The Worldwide LHC Computing Grid (WLCG) provides global computing resources for the storage, distribution, and analysis of data generated by the Large Hadron Collider (LHC). Combining around 1.3 million CPU cores across 164 data centers worldwide, the WLCG is a cornerstone of modern high-energy physics research. As computational demand increases, CERN needs reliable accounting of the computing resources. Accurate accounting enables experiments and stakeholders to understand available computing resources and utilize them effectively.

The High Energy Physics (HEP) community has agreed to use dedicated benchmarks for reporting computational capacities. In April 2023, the previously used HEP-SPEC06 (HS06) \cite{physics:hs06} benchmark was replaced by HEPScore23 (HS23) \cite{chep:hs23}. As part of this transition, it became essential to compare the CPU corepower, which is performance score per CPU core, published by ATLAS \cite{experiment:atlas} sites with the runtime corepower observed while running regular jobs in production environment. One of the primary aims was to verify the conversion rate between the old and new benchmark scores and ensure consistency across sites. Traditionally, sites report their performance values annually, providing weighted averages of different corepower values from various CPU models available at this site. However, until now, there was no direct method to validate the numbers provided by these sites. The infrastructure described in this work is designed to monitor and validate the reported values in a production environment by running the benchmark on the grid. This approach not only verifies the annual performance numbers but also enhances transparency and reliability in resource accounting. It also allow for early detection of incorrect values and fix them as soon as it is possible. 

\section{Submission Infrastructure}
\label{section:infra}
\begin{figure}[ht]
    \centering
    \includegraphics[width=1\linewidth]{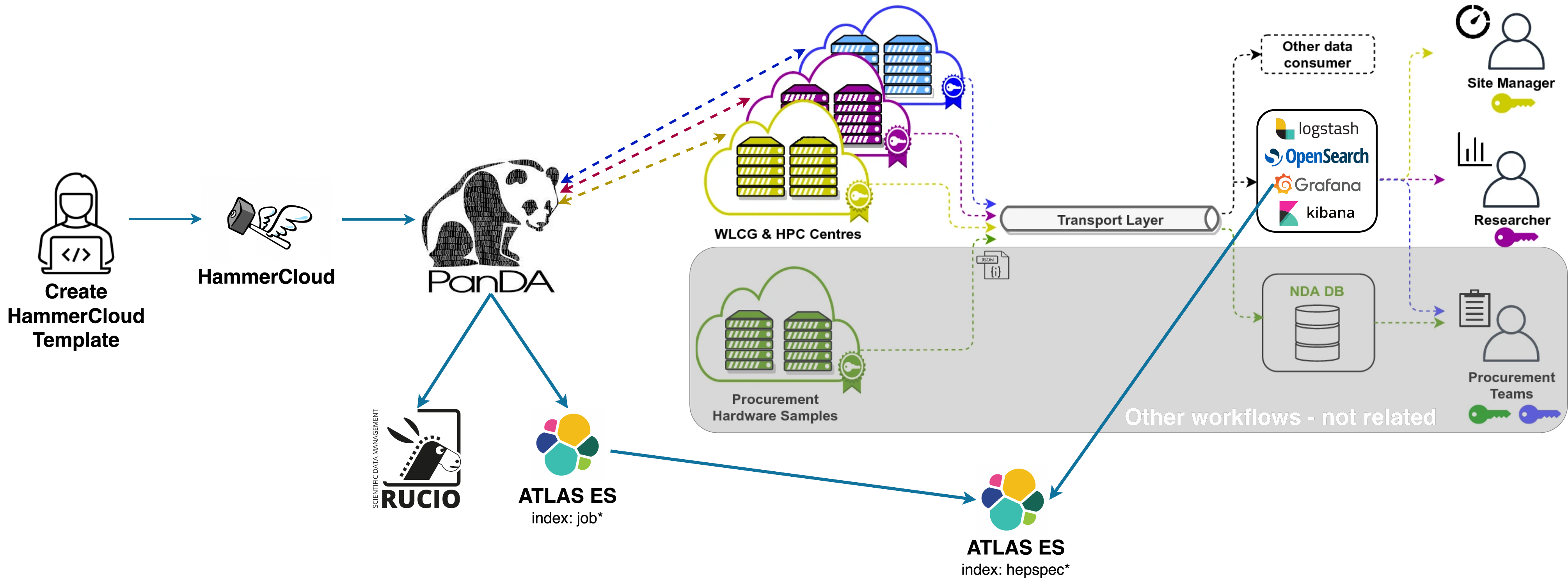}
    \caption{Submission infrastructure includes HammerCloud, PanDA, HEP Benchmark Suite, ActiveMQ, OpenSearch, Grafana, Elasticsearch and Kibana enabling automatic and continuous measurements of runtime corepower of compute resources provided by the WLCG sites to ATLAS.}
    \label{fig:infra}
\end{figure}
The submission infrastructure presented in \textbf{Figure~\ref{fig:infra}} is an ATLAS adaptation of the workflow described in detail in \cite{acat2024benchmarking}. HammerCloud (HC) \cite{web-hammercloud}, a framework used for testing and benchmarking WLCG resources, is used to dispatch identical jobs at regular intervals via PanDa \cite{atlas:panda}. To achieve this, a HC template defining the workload, the environment, the submission schedule, and the list of targeted queues is created. \textit{PanDA queue} is a concept on the PanDA Workflow Management System (WMFS) of grouping and configuring a set of resources together needed for processing workflows. Each WLCG Site can include multiple PanDA queues. Every four hours, the job is submitted at each queue. The job runs the enhanced HEP Benchmark Suite \cite{web:hepsuite}, executing the HEPScore23 configuration, including 7 workloads from 5 experiments. Upon job completion, the benchmarking results are sent through ActiveMQ \cite{web:activemq} to the benchmarking OpenSearch \cite{web:opensearch} instance and are available for inspection on a dedicated benchmarking Grafana \cite{grafana} dashboard; simultaneously, PanDA-specific information is sent to the ATLAS Elasticsearch (ES) \cite{web:elasticsearch} instance. Subsequently, data from both sources is combined using a Python script and transmitted to the ATLAS ES instance under the \texttt{hepspec*} index. The ATLAS ES instance hosts a Kibana \cite{web:kibana} dashboard for comprehensive visualization and analysis of the combined data.

\section{Data Analysis}
\label{section:data-analysis}
Over the past year, a large amount of data has been collected using the infrastructure described in Section~\ref{section:infra}. \textbf{Table~\ref{tab:statistics}} provides a statistical summary of the collected data. Each job runs on 8 cores, following the WLCG standards \cite{web:wlcg-standards}. In each run, performance metrics of each server were collected, along with additional system information, including user-defined metrics such as machine load during script execution, average CPU frequency, memory and swap. These extensive statistics allowed for a detailed analysis of the system performance observed during benchmark runs and enabled comparison with the numbers reported by the sites.
\begin{table}[hb]
  \centering
  \caption{Benchmarking jobs statistics}
\begin{tabular}{c c c c c c}
    \toprule
    \textbf{No. \#} & \textbf{No. Sites} & \textbf{No. CPU} & \textbf{Walltime [min]} & \textbf{\% of total walltime} \\
    \midrule
    187045 & 139 & 251 & 81 & 0.06\\
    \bottomrule
  \end{tabular}
  \label{tab:statistics}
\end{table}

\subsection{Declared and Runtime Corepower}
\label{subsection:corepower}
The corepower of a server has historically been defined as the HS06 per core, serving as a key metric to evaluate computing capabilities based on specific hardware. With the transition from HS06 to HS23 in April 2023, the corepower metric is now represented as HS23 per core. To ensure a smooth transition, it was initially agreed that corepower values would be converted one-to-one for each site. Currently, HS23 is the WLCG standard, and sites are expected to provide updated corepower values.

The corepower value reported by a site, referred to here as \textit{declared corepower}, represents the weighted average of the corepower of different CPU models available at given queue on a site. A single site may consist various hardware resources that are grouped into different queues. The objective of this study was to compare the declared corepower values in ATLAS-CRIC \cite{web:atlas-cric} with the performance observed during job execution, hereafter referred to as the \textit{runtime corepower}. To calculate runtime corepower, the weights of different CPU models available at each queue were first derived using all available data from PanDA jobs, leveraging the \texttt{walltime\_x\_core}, defined as the product of job walltime and the number of cores. The weight of each CPU model (x) at a given queue was computed as defined in \textbf{\ref{eq:weight}}. 
\begin{equation}
w_x = \frac{\sum_{i \text{ on jobs}} \text{walltime\_x\_core}_i^x}{\sum_{x \text{ on CPU}} \sum_{i \text{ on jobs}} \text{walltime\_x\_core}_i^x}
\label{eq:weight}
\end{equation}
Subsequently, the weighted average corepower for each queue was calculated using only CPU models for which benchmarking data was available, as presented in \textbf{\ref{eq:corepower_runtime}}. 

\begin{equation}
\text{corepower\_runtime}^{\text{queue}} = 
\frac{\sum_{x \text{ on CPU}} w_x \cdot \text{corepower\_runtime}_x^{\text{queue}}}{\sum_{x \text{ on CPU}} w_x}
\label{eq:corepower_runtime}
\end{equation}

For the analysis, only queues with complete weights data were included. The \textit{relative change} in corepower for a given queue (q) was then defined as the ratio of runtime corepower to declared corepower, minus one, as presented in \textbf{\ref{eq:relative_change}}.

\begin{equation}
\text{relative change} = \frac{\text{corepower\_runtime}_q}{\text{corepower\_declared}_q} - 1
\label{eq:relative_change}
\end{equation}

It is important to note that systematic uncertainties are inherent in the measurements, and these were carefully considered in this study. Sources of uncertainty include variations in HS23/HS06 scaling \cite{springer:2021:misc}, fluctuations in runtime HS23 probes, dynamic CPU configurations as switching between hyper-threading enabled and disabled without reporting, performance variations due to system load, and errors in weight calculations critical to the analysis. To address these uncertainties, a discrepancy threshold of ±25\% was established. Only relative changes exceeding this threshold were classified as significant discrepancies, so those ones where numbers reported by queues differ substantially from measured runtime corepower.

\section{Analysis of Relative Change Corepowers}
\label{section:relative-change}

\begin{figure}[ht]
    \centering
    \includegraphics[width=1\linewidth]{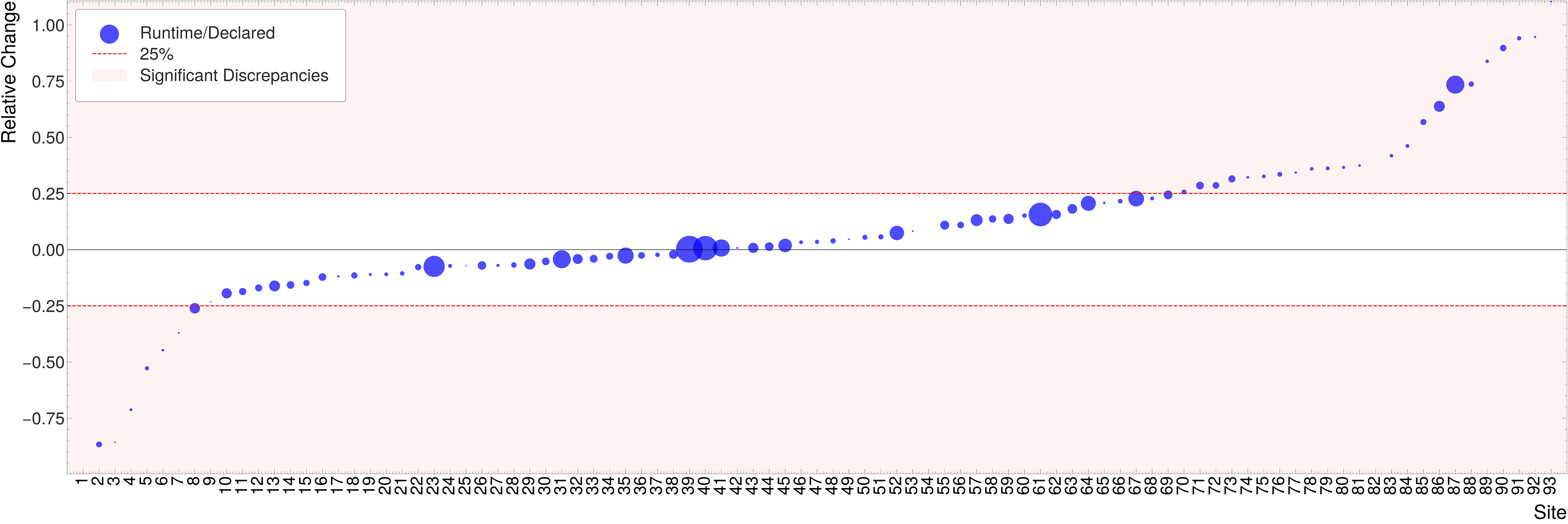}
    \caption{Relative Change following \textbf{\ref{eq:relative_change}} for PanDA queue, grouped by site. More than one data point per site indicates that the site provides more than one PanDa queue. The shadowed region highlights critical discrepancies. The size of the marker is proportional to site contribution level calculated based on its walltime\_x\_core.}
    \label{fig:relativechange}
\end{figure}

Measurements carried out throughout 139 sites enabled a comprehensive corepower analysis for 72 of them, where complete weight data was available, as shown in \textbf{Figure~\ref{fig:relativechange}}. The different sites were anonymised and shown on the \texttt{x-axis}, with the relative change on the \texttt{y-axis}. The conservative acceptance threshold of 25\% highlights the large discrepancies. A relative change of zero represents a perfect match between the declared and the runtime corepower, it can be seen that most of the points are within the treshold range. Positive relative change indicate underreporting of the declared corepower by the queue on given site, whereas negative values suggest overreporting. From \textbf{Figure~\ref{fig:relativechange}}, it can be observed that 32\% of the analyzed sites show critical discrepancies. Two hypotheses were considered to explain these discrepancies: (a) the queues at these sites are either underloaded or overloaded, and/or (b) the declared corepower values are inaccurate.

\subsection{Corepower vs Load}
\label{subsection:corepower-vs-load}
To test the first hypothesis, a corepower versus load analysis was conducted. The average load information during the benchmark execution was retrieved and correlated with the measured performance. 
\begin{figure}[ht]
    \centering
    \begin{subfigure}[t]{0.47\textwidth}
        \includegraphics[width=\linewidth]{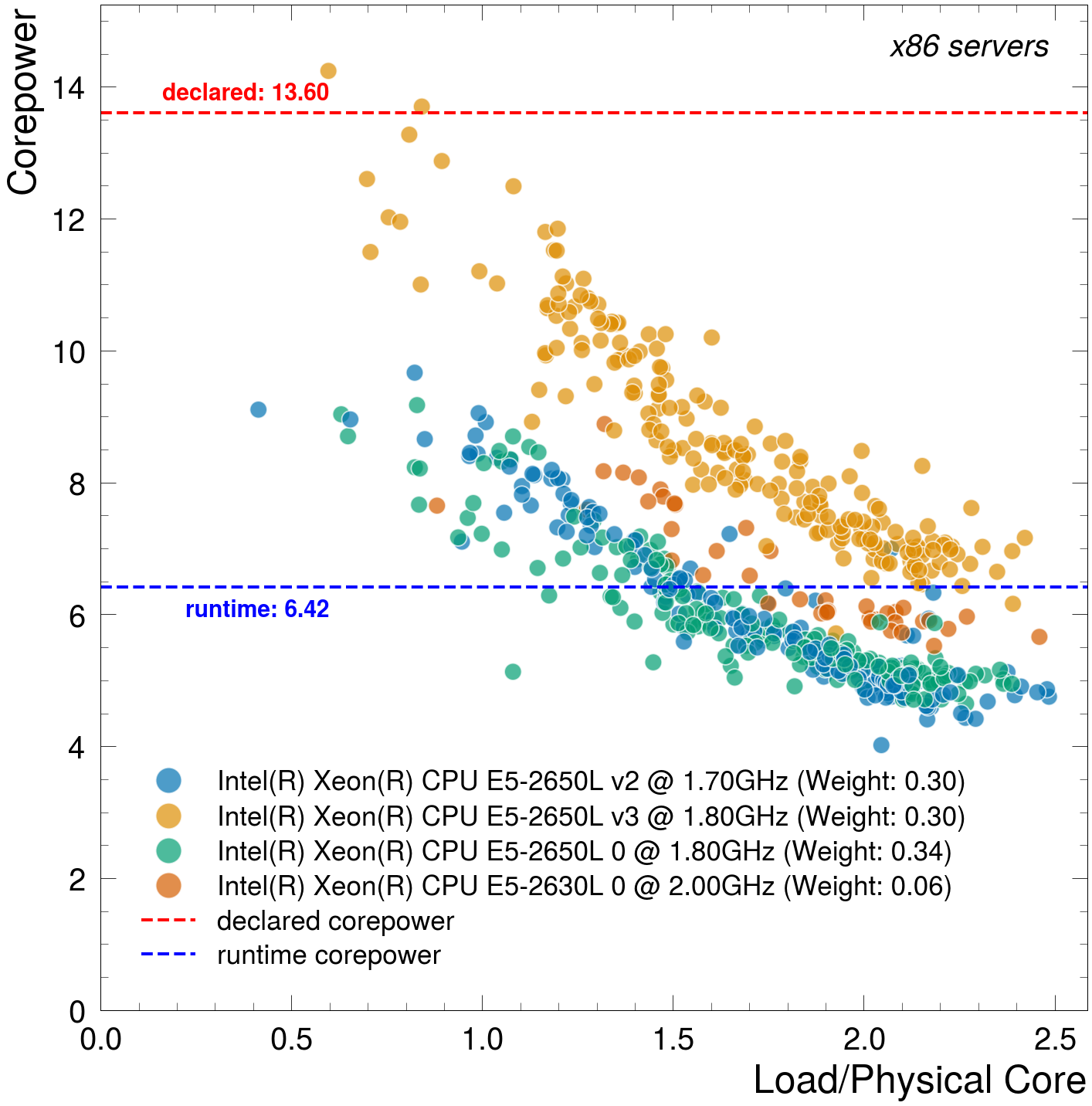}
        \caption{Queue with negative relative change}
        \label{fig:site-a}
    \end{subfigure}
    \hfill
    \begin{subfigure}[t]{0.47\textwidth}
        \includegraphics[width=\linewidth]{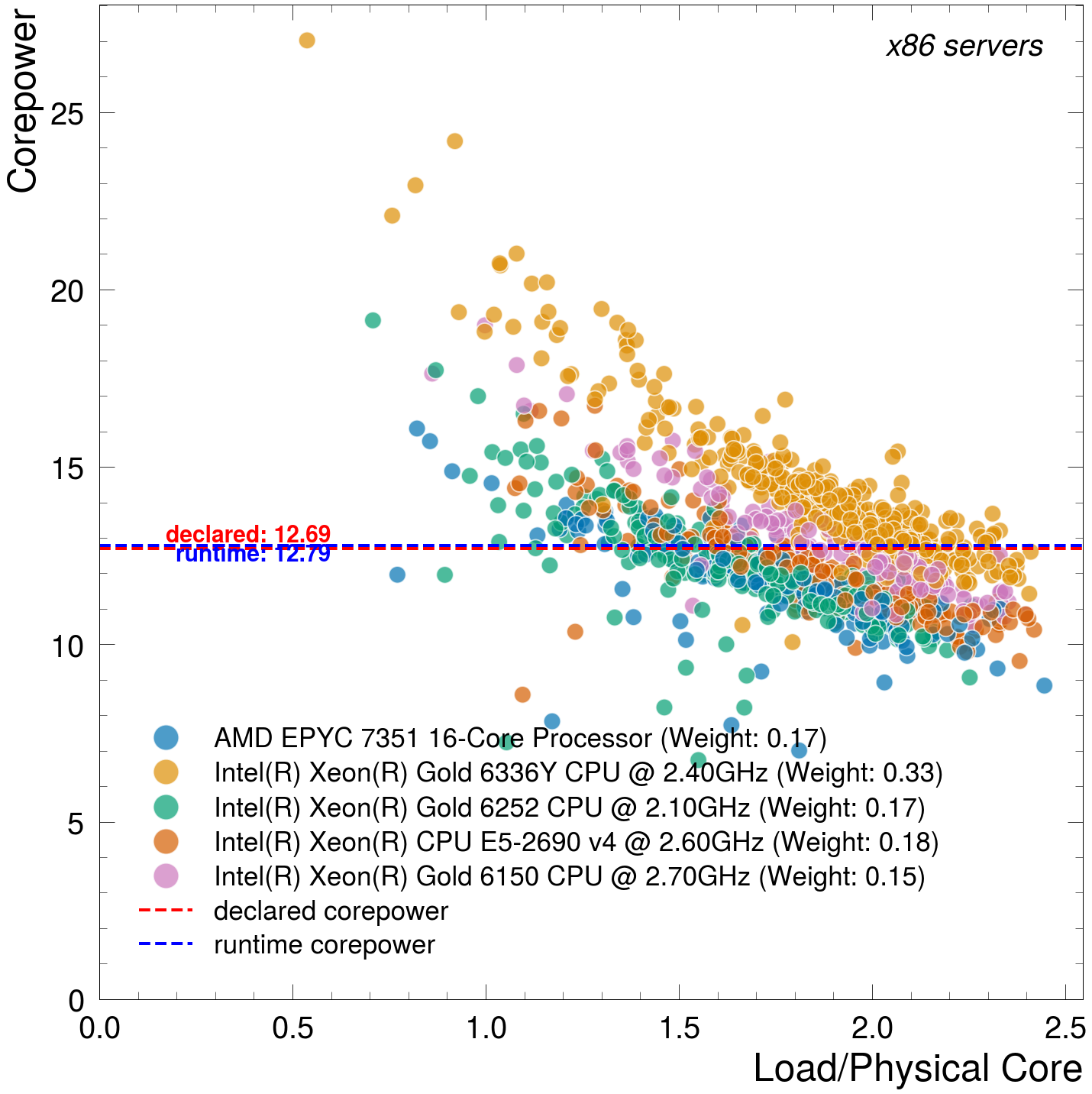}
        \caption{Queue with neutral relative change}
        \label{fig:site-b}
    \end{subfigure}
    \vskip\baselineskip
    \begin{subfigure}[t]{0.47\textwidth}
        \includegraphics[width=\linewidth]{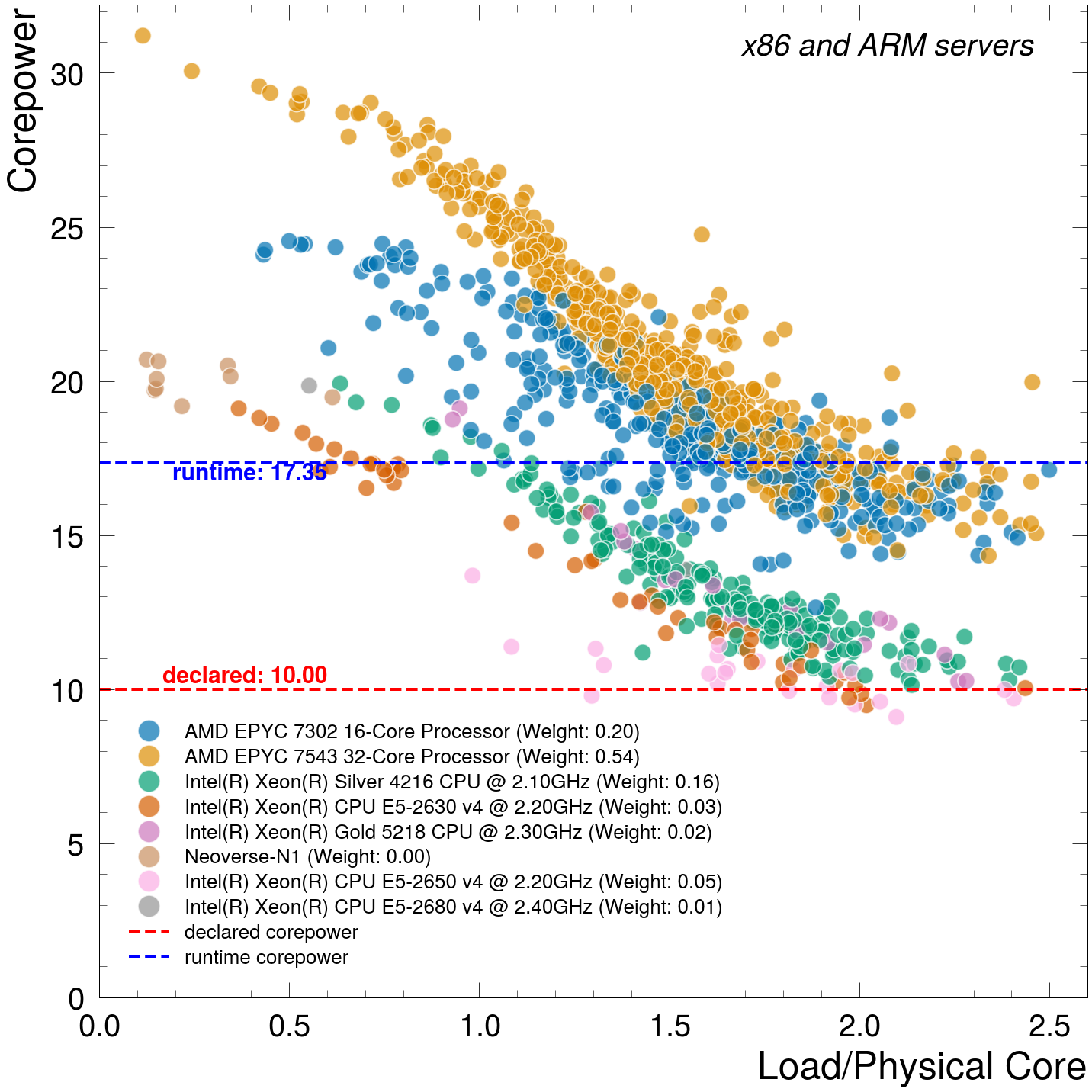}
        \caption{Queue with positive relative change}
        \label{fig:site-c}
    \end{subfigure}
    \hfill
    \begin{subfigure}[t]{0.47\textwidth}
        \includegraphics[width=\linewidth]{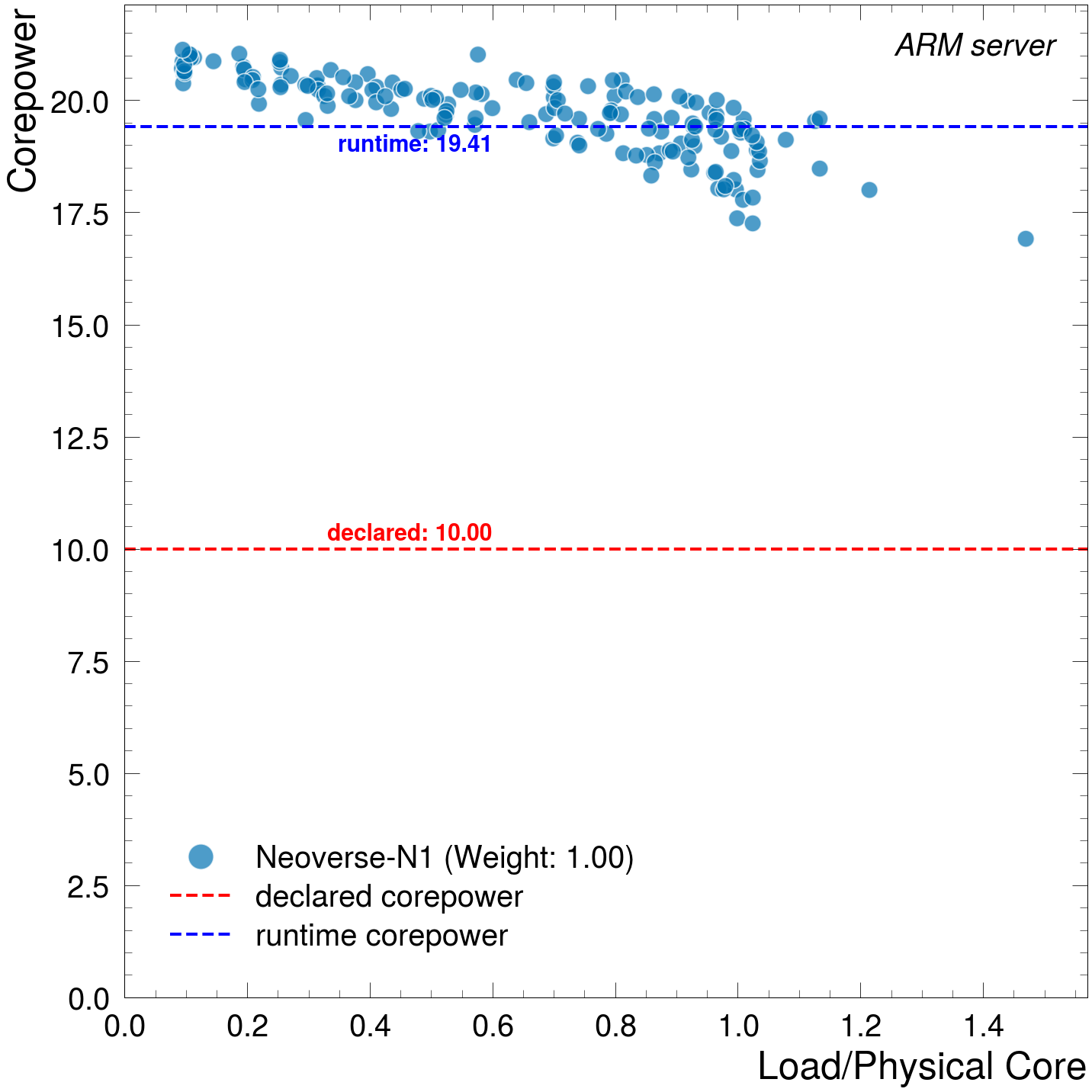}
        \caption{Queue with positive relative change}
        \label{fig:site-d}
    \end{subfigure}
    \caption{Corepower vs load/core correlation of 4 typical cases of queues with: (a) negative relative change, (b) neutral relative change, (c) positive relative change, (d) positive relative change, ARM server. The dashed red and blue lines represent the declared and runtime corepowers respectively. Each color marker identifies a different CPU model on a given queue.}
    \label{fig:corepower-vs-load}
\end{figure}
The load/physical core metric is defined as the measured load divided by the number of physical cores. As a consequence a fully loaded server with hyper-threading disable or enable will have a load/physical core value of 1 or 2 respectively. In \textbf{Figure~\ref{fig:corepower-vs-load}}, the plot reveals a correlation between machine load/physical core and performance: as the load increases, the performance decreases. This is being especially visible in the range from 1 to 2 of load/physical core.

Four examples were selected to illustrate typical behaviors with respect to the relative change results. \textbf{Figure~\ref{fig:site-a}} shows a queue where the declared corepower exceeds the runtime corepower. This queue was fully utilized, but the CPU models in use are from 2012. In this case, the scaling ratio between HS23 and HS06 is 0.7, as taken from Figure 4 in \cite{chep:hs23}, which explains the detected negative relative change. This shows that the values for very old queues should be updated as soon as possible to better plan resource utilization. The second case, shown in \textbf{Figure~\ref{fig:site-b}}, represents one of the largest queues, where the relative change is close to zero. Here, the site demonstrates efficient utilization across all available servers, with no discrepancies observed. 

The \textbf{Figure~\ref{fig:site-c}} provides a third example suggesting that, in many cases, the server is not fully loaded, which could be one of the reasons for the observed higher runtime corepower compared to the declared value. However, this does not appear to be the sole cause. As seen in the figure, there are two servers whose performance match precisely the declared corepower value of 8.74. This observation suggests that the declared corepower might not have been updated to reflect the presence of other, more modern servers in the queue. Instead, it appears to rely on outdated values associated with older hardware, failing to account for the contribution of additional servers with higher performance capabilities.

The fourth example, presented in \textbf{Figure~\ref{fig:site-d}}, is a special case involving a single CPU model on a given queue, Neoverse-N1 based on ARM architecture. Unlike the x86 servers, weaker correlation between load and performance is observed. However, a significant discrepancy is evident — a factor of nearly two between the runtime corepower and the declared corepower, favoring the runtime measurement.

The analysis shows that even if there is correlation between the average load of a server and its performance, it is not the only cause of the discrepancies. But it is important to notice that if all servers in a queue would have been consistently underloaded, a positive relative change would be observed, meaning the runtime corepower would appear higher than the declared value. On the other hand, if all servers in a queue would have been consistently overloaded, what is defined as a load above 2 for systems with hyper-threading enabled or above 1 for those with hyper-threading disabled — a negative relative change would be observed, indicating the runtime corepower would appear lower than the declared value, if the declared value would be set correctly.

\begin{figure}[h!]
    \centering
    \includegraphics[width=1\linewidth]{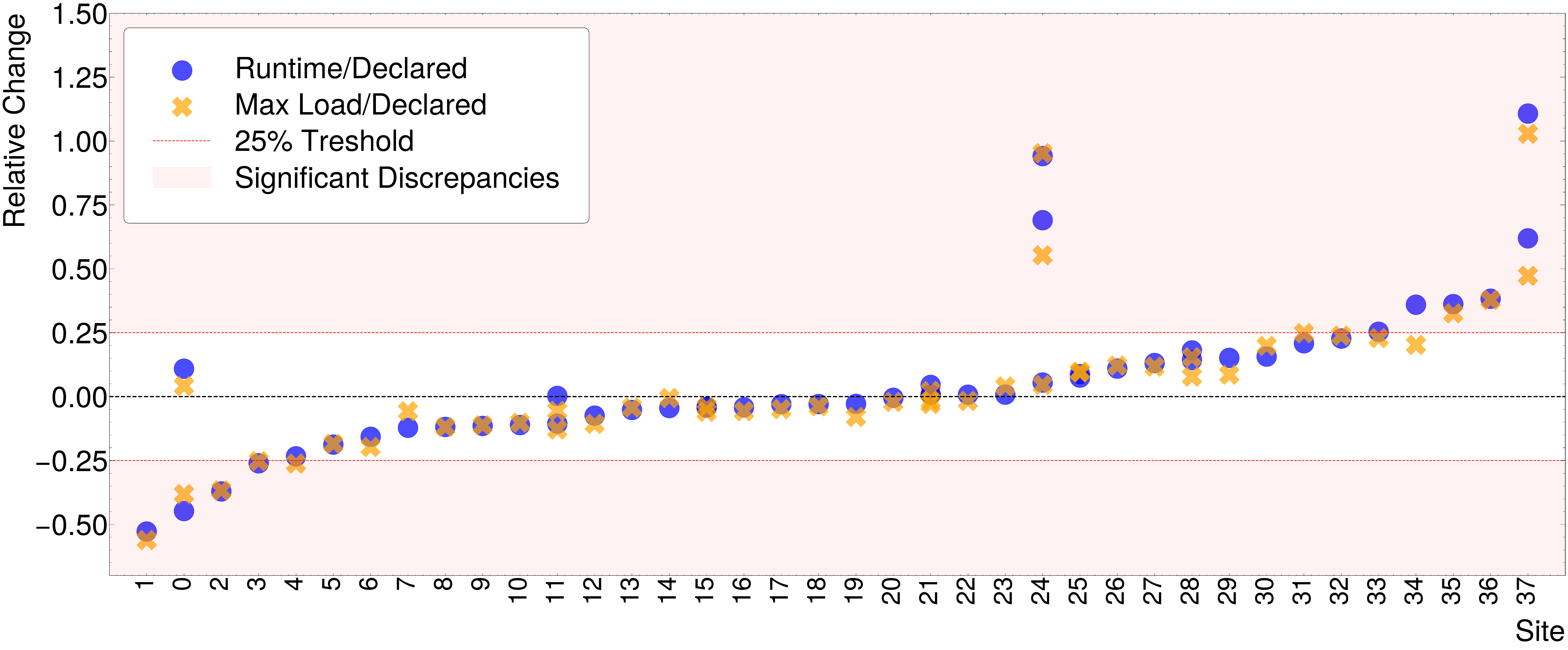}
    \caption{Relative change for different sites calculated when servers were fully loaded}
    \label{fig:relative-loaded}
\end{figure}

Another consideration related to the load involves the method of calculating runtime corepower, which is based on the average performance across the entire load range. It was important to verify whether this approach could be responsible for the observed behavior of different queues. To address this, a separate analysis was conducted using only measurements taken when servers were fully loaded. It was assumed that this approach would reduce discrepancies in relative change to within the established threshold.

The results, presented in \textbf{Figure~\ref{fig:relative-loaded}}, show fewer data points, as fewer servers were fully loaded during the measurements. While this approach did slightly reduce discrepancies in some of the cases, it did not bring them entirely within the expected threshold. This finding confirms that calculating performance across the full load range is valid and that the uncertainty associated with this method is minimal and acceptable for the analysis.

\subsection{Analysis of Declared Corepower Data Sources}
\label{subsection:data-source}

The declared corepower values used by ATLAS are sourced from the ATLAS-CRIC, where they are directly being reported by site administrators from each site or loaded from other data sources. A detailed analysis of these values revealed that 80\% of the PanDA queues were cloned from pre-existing ones, with 50\% inheriting their corepower values from the original queues. This practice has significant implications for multiple queues employing newer architectures, such as x86 and ARM-based queues shown in \textbf{Figure~\ref{fig:site-c}} and \textbf{Figure~\ref{fig:site-d}}, where the discrepancy with the outdated declared corepower is larger. 

ATLAS-CRIC serves as the central place, where site administrators can update these values, but it is also an integration point for multiple data sources. During this analysis, discrepancies between the various data sources were also identified and reported accordingly, with corrective actions currently underway.

Considering the combined contributions from all sites and weighting them by their \texttt{walltime\_x\_core}, the discrepancies between the declared corepower and the runtime corepower values result in an overall 6\% advantage in favor of the runtime performance. While this result is positive, it highlights the need to update declared corepower values. Failure to do so could increasingly affect decision-making processes that rely on accurate reported data.

\section{Conclusions}
\label{section:conclusions}
The automated HEPScore23 submission infrastructure via PanDA and HammerCloud provides a novel approach, previously unavailable for validating the corepower values declared by sites, offering a reliable cross-check for the official accounting system. By leveraging the HEP Benchmark Suite, additional system metrics were collected, enabling a detailed analysis of the runtime versus the declared corepower values. The observed relative changes revealed significant discrepancies for several queues, highlighting differences between declared and actual runtime performance. These discrepancies have the potential to impact decision-making processes, leading to inefficiencies and complications in resource allocation.

A systematic analysis has shown that outdated declared corepower values are a primary source of these discrepancies, underlining the importance of regular updates in the data source. While server load does influence performance, it alone does not fully justify the observed differences. The presented methodology for measuring the runtime corepower proved effective in identifying queues on sites with large discrepancies, enabling quick reporting and resolution. This approach not only ensures better accuracy and transparency in resource accounting but also enhances operational efficiency and fosters better decision-making for the ATLAS experiment and WLCG at large.
%
%
%
\bibliography{CHEP.bib}

\end{document}